\documentclass[]{elsarticle}
\usepackage{natbib}

\usepackage{geometry}
\usepackage[]{fontenc}
\usepackage[latin1]{inputenc}
\usepackage{listings}

\usepackage{natbib}
\usepackage{amsmath}
\usepackage{graphicx}
\usepackage{amssymb}

\def\apgt{\ {\raise-.5ex\hbox{$\buildrel>\over\sim$}}\ }
\def\aplt{\ {\raise-.5ex\hbox{$\buildrel<\over\sim$}}\ }
\def\lt{\ {\raise-.5ex\hbox{$\buildrel>$}}\ }
\def\gt{\ {\raise-.5ex\hbox{$\buildrel<$}}\ }

\title{SAPPORO: A way to turn your graphics cards into a GRAPE-6}

\author[a1,a2,a3]{Evghenii Gaburov\corref{fn1}}
\author[a2,a3]{Stefan Harfst}
\author[a2,a3,a1]{Simon Portegies Zwart}
\address[a2]{Astronomical Institute ``Anton Pannekoek'', University of Amsterdam}
\address[a3]{Section Computational Science, University of Amsterdam}
\address[a1]{Leiden Observatory, University of Leiden\\
the Netherlands}
\cortext[fn1]{egaburov@strw.leidenuniv.nl}

\begin{document}

\begin{abstract}
  We present {\tt Sapporo}, a library for performing high-precision
  gravitational $N$-body simulations on NVIDIA Graphical Processing
  Units (GPUs). Our library mimics the GRAPE-6 library, and $N$-body
  codes currently running on GRAPE-6 can switch to {\tt Sapporo} by a
  simple relinking of the library.  The precision of our library is
  comparable to that of GRAPE-6, even though internally the GPU
  hardware is limited to single precision arithmetics. This limitation
  is effectively overcome by emulating double precision for
  calculating the distance between particles. The performance loss of
  this operation is small ($\aplt 20$\%) compared to the advantage of
  being able to run at high precision. We tested the library using
  several GRAPE-6-enabled N-body codes, in particular with {\tt
    Starlab} and {\tt phiGRAPE}.  We measured peak performance of
  800\,Gflop/s for running with $10^6$ particles on a PC with four
  commercial G92 architecture GPUs (two GeForce 9800GX2). As a
  production test, we simulated a 32k Plummer model with equal mass
  stars well beyond core collapse. The simulation took 41 days, during
  which the mean performance was 113\,Gflop/s. The GPU did not show
  any problems from running in a production environment for such an
  extended period of time.

\end{abstract}
\maketitle

\section{Introduction}
 
Graphical processing units (GPUs) are quickly becoming main stream in
computational science. The introduction of Compute Unified Device
Architecture \citep[CUDA,][]{GPUGems1}, in which GPUs can be
programmed effectively, has generated a paradigm shift in scientific
computing \citep{2007astro.ph..3485H}. Modern GPUs are greener in
terms of CO$_2$ production, have a smaller footprint, are cheaper, and
as easy to program as traditional parallel computers. In addition, you
will not have a waiting queue when running large simulations on your
local GPU-equipped workstation.

Newtonian stellar dynamics is traditionally on the forefront of
high-performance computing.  The first dedicated Newtonian solver
\citep{1986LNP...267...86A} was used to study the stability of the
solar system \citep{1992Sci...257...56S}. And soon even faster
specialized hardware was introduced by the inauguration of the GRAPE
family of computers, which have an impressive history of breaking
computing speed records \citep{1998sssp.book.....M}.

Nowadays, the GPUs are being used in various scientific areas, such as
molecular dynamics \citep{Anderson20085342, citeulike:3156232},
solving Kepler's equations \citep{2009NewA...14..406F} and Newtonian
$N$-body simulations. Solving the Newtonian $N$-body problem with GPUs
started in the early 2000s by adopting a shared time step algorithm
with a 2nd order integrator \citep{Nyland04}. A few years later this
algorithm was improved to include individual time steps and a higher
order integrator \citep{2007NewA...12..641P}, in a code that was
written in the device specific language Cg \citep{Cg}. The performance
was still relatively low compared to later implementations in CUDA via
the {\tt Cunbody} package \citep{2007astro.ph..3100H}, {\tt Kirin}
library \citep{2008NewA...13..103B}, and the {\tt Yebisu} $N$-body
code \citep{2008NewA...13..498N, 2008PHD_NITADORI}. The main problem
with the two former implementations was the limited accuracy of the
GPU, which only enabled single precision.  In part this problem was
solved with the introduction of the double precision GPU (GTX280), but
at a dramatic performance-hit as only a limited number of processor
pipelines supported double-precision calculations\footnote{\tt
  http://www.nvidia.com/object/product\_tesla\_s1070\_us.html}.

In this paper we introduce {\tt Sapporo}\footnote{The {\tt Sapporo}
  library is free and can be downloaded from {\tt
    http://modesta.science.uva.nl}.}, a library, which is written in
CUDA, for running gravitational $N$-body simulations on NVIDIA GPUs. The
{\tt Sapporo} library closely matches the calling sequence of the
GRAPE-6 library \citep{2003PASJ...55.1163M}, and codes which are
already running on a GRAPE-6 can be immediately be run on GPUs without
any modifications to their source code.

Here, we describe the implementation of our library, and present both
accuracy and performance measurements using {\tt Sapporo}. For the
latter, we use two direct $N$-body simulations environment, one being
{\tt Starlab} \citep{2001MNRAS.321..199P} and the other is {\tt
  phiGRAPE} \citep{2007NewA...12..357H} as it is implemented in the
Multi Scale Software Environment \citep[MUSE,][]{2009NewA...14..369P}.

Our new implementation of the $N$-body force calculation in CUDA is an
important step towards high-precision direct $N$-body simulations using
GPUs, as the library is more flexible and more general than previous
implementations. In addition we identify the operations in the code
where double precision accuracy is most important and implement double
precision arithmetic in those locations. The cost for doing this is
limited to a $\sim 20$\% loss in performance (but overall performance
is still very high).

\section{Implementation}\label{sect:methods}

We have designed a library to calculate gravitational forces on a GPU
in order to accelerate $N$-body simulations. The library can be used
in combination with a standard 4$^{\rm th}$-order predictor-corrector
Hermite integration scheme (Makino \& Aarseth 1992), either with
shared or block time steps. Such a scheme consists of three essential
steps: predictor, force calculation, and corrector. In the predictor
step, the positions and velocities of all particles are predicted for
the next time step. Then, the gravitational accelerations and their
first derivatives (jerks) are calculated using the predicted positions
and velocities. Finally, the predicted positions and velocities are
corrected using the newly computed accelerations and jerks. In the
case of a block time step scheme, the last two steps are only executed
on a block of active particles. In the following, we will focus on
block time steps, since shared time step scheme can be considered a
variant of the block time step scheme.

In the case of the block time steps, the system is divided into
$i$-particles and $j$-particles, which are the sinks and sources of
the gravitational forces\footnote{In the following, by gravitational
  forces we imply gravitational accelerations, potentials and jerks.},
respectively. Within the Hermite scheme, we take the following
subsequent steps: we predict the positions and velocities of the $i$-
and $j$-particles (predictor step); calculate the gravitational forces
exerted by the $j$-particles on the $i$-particles (calculation step);
correct the positions and velocities of the $i$-particles (corrector
step).

The actual parallelisation of the force calculation is not too
difficult and many examples for parallel $N$-body algorithms
exist \citep{journals/pc/GualandrisZT07, 775815}. Implementing the
algorithm on a GPU is a little more challenging due to the specific
design of GPUs. For example, a G80/G92 NVIDIA GPU consist of 16
multi-processors (MPs) each being able to execute up to 768 threads in
parallel. A parallel, SIMD (single instruction multiple data), element
on such a GPU is called a warp. A warp is a set of 32 threads which
execute the same instruction but operate on different data, and up to
24 warps can be executed in parallel on each MP. In total, a single
G80/G92 GPU is able to execute up to 12288 threads in parallel, which
means that a program which runs on a GPU, called a kernel, should be
able to efficiently exploit such a high degree of parallelism.

In addition, the GPU design impose a strict memory access pattern on
such a kernel. For example, a GPU has two types of memory: global
memory, which is an equivalent of RAM on the CPU, and the shared
memory, which is equivalent to L1-cache on the CPU. The global memory
is further subdivided into local memory which is allocated on a per
thread basis, and both texture and constant cached read-only
memory. The major difference between the shared and the global memory
is the access latency. Access to an element of the shared memory is as
fast as access to a register, whereas access to an element in the
global memory has a latency of 400-600 cycles. Since the data
initially reside in the global memory, each of the parallel thread
should cooperatively load the data to the shared memory in a specific
pattern in order to reduce the access latency. The threads can then
efficiently operate on the data in the shared memory.

\subsection{Task decomposition between the CPU and the GPU}

At first, we estimate the complexity of each of the steps in the
Hermite scheme. We assume that the number of $j$-particles is equal to
$n$, and the number of $i$-particles is equal to $m$. The time
complexities of the predictor, the calculation, and the corrector
steps are ${\cal O}(n)$, ${\cal O}(nm)$, and ${\cal O}(m)$,
respectively.  It is important to note, that the $n^2$-scaling of the
force calculation has been reduced to $nm$ by the introduction of
block time steps. As a result, the time contribution of the
calculation and of the corrector steps decreases with block size,
whereas that of the prediction step remains constant. This is
important as block sizes are typically small on average in direct
$N$-body simulations.

\begin{figure}
  \begin{center}
    \includegraphics[scale=0.5]{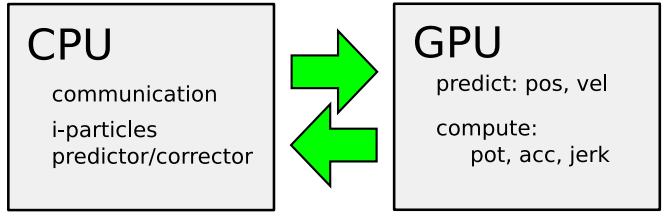}
  \end{center}
  \caption{Decomposition of the tasks between the CPU and the GPU. The
    CPU is responsible for the communication, prediction of the
    $i$-particles and the correction step, whereas the prediction of
    the $j$-particles and the force calculation step are carried
    out on the GPU.}
  \label{fig:decompose}
\end{figure}

Motivated by this, we implemented both the prediction of $j$-particles
and the force calculation on the GPU. The prediction of $i$-particles
and the corrector step are carried out on the CPU
(Fig. \ref{fig:decompose}); in fact, GRAPE-enabled $N$-body codes have
exactly the same decomposition. Another motivation for this
decomposition is the communication overhead: if the predictor step
would be carried out on the CPU, all $n$ particles would have to be
communicated to the GPU at every time-step. In this case,
communication would become a bottle-neck for the calculation.  In our
chosen implementation, only $i$-particles need to be communicated, and
this therefore decreases the communication overhead.

\subsection{Predictor step}

In this step, the GPU's task is to predict positions and velocities of
the $j$-particles. It is a rather trivial step to parallelise since a
particle does not require information from other particles. Hence, we
assign a particle to each of the parallel threads on the GPU. Each
thread reads for its particle position (${\bf x}$), velocity (${\bf
  v}$), acceleration (${\bf a}$), jerk (${\bf j}$), and time step
($dt$) from global memory into registers. This data is required to
execute the predictor step:
\begin{eqnarray}
  {\bf x}_{\rm pred} &=& {\bf x}_0 + {\bf v}_0 dt + {\bf
    a}_0\frac{dt^2}{2} + {\bf j}_0 \frac{dt^3}{6}, \label{eq:Ppos}
  \\ {\bf v}_{\rm pred} &=& {\bf v}_0 + {\bf a}_0 dt + {\bf
    j}_0\frac{dt^2}{2}, \label{eq:Pvel}
\end{eqnarray}
where subscripts ``$0$'' and ``${\rm pred}$'' refer to the initial and
predicted values, respectively; both the predicted position and
velocity are stored in the global memory for later use. The prediction
of the $i$-particles is similarly implemented on the CPU.

Prediction of particle positions must be computed in double
precision (DP) arithmetics because the integration scheme is most
sensitive to round-off errors in positions
\citep{1985mts..conf..377A}. This creates difficulties for
implementing the predictor on the GPU due to the following reason:
previous generations of NVIDIA GPUs (G80/G92) do not natively support
DP arithmetics, and the current generation (GT200) executes DP
instructions an order of magnitude slower than their single precision
(SP) equivalents. However, not all operations in Eq.~\ref{eq:Ppos}
require double precision. Only the position is stored in DP, whereas
the rest of the terms are stored in SP. Consequently, the
multiplications and divisions can be carried out in SP, but the
summation should be carried out in DP. This can be done most
efficiently by emulating DP only where it is necessary, instead of
implementing all of Eq.~\ref{eq:Ppos} in DP. As a result, the loss in
performance is acceptable and the implementation is not dependent on
hardware supporting DP arithmetics.

The memory storage requirement for DP float (64 bit) is twice that of
a SP float (32 bit). Therefore, it is natural to store DP numbers in
two SP numbers: one of the SP numbers stores the most significant
digits, the other stores the least significant digits. Such a
representation of DP is known as a double-single (DS)\footnote{\tt
  http://crd.lbl.gov/$\sim$dhbailey/mpdist/}. In this representation,
the number of significant digits is 14 compared to 16 in
DP\footnote{The mantissa of DP consists of 53 bits, compared to 48
  bits in DS because a SP number has a mantissa of 24 bits.}, but
nevertheless it is a factor of two larger than in SP.

In the following, we will use DS to emulate DP where it is
required. In Eq.~\ref{eq:Ppos}, this only needs to be done for the
positions ${\bf x}_{0}$ and ${\bf x}_{\rm P}$. All multiplications and
divisions are carried out and have their results stored in SP. These
SP numbers are then added together and the resulting number is added
to the DS number\footnote{Addition of DS with SP operation requires 10
  SP floating point operations (FLOP)} ${\bf x}_{0}$. The result is
stored again in DS as ${\bf x}_{\rm P}$. Therefore, the floating point
operation (FLOP) count for Eq.~\ref{eq:Ppos} is 1\,DS + 10\,SP
operations or 11 FLOP per particle; the Eq.~\ref{eq:Pvel} requires
only 6 FLOP per particle.

\subsection{The calculation step}

\begin{figure}
  \begin{center}
    \includegraphics[scale=0.75]{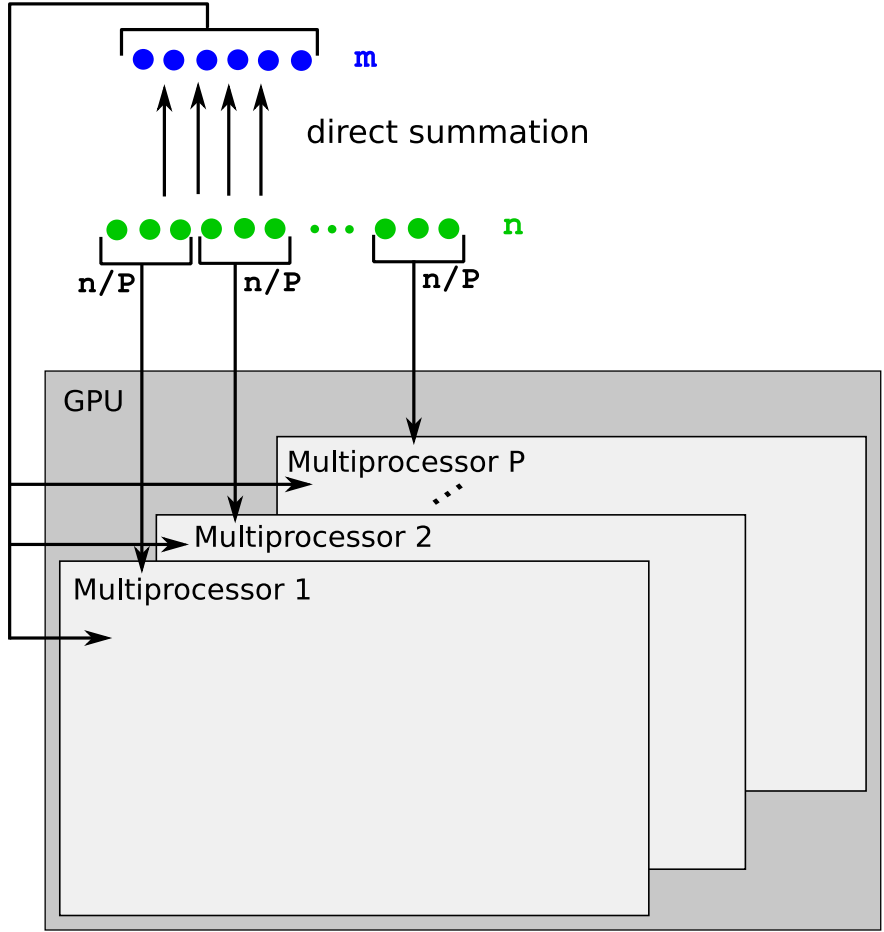}
  \end{center}
  \caption{An illustration of the decomposition between
    multiprocessors (MP) on the GPU. The $n$ $j$-particles are equally
    distributed between all $P$ MPs, where $P$ is the number of
    MPs. An identical copy of all $m$ $i$-particles is distributed to
    each of the MPs. As a result, every MP computes gravitational
    forces from $n/P$ bodies on the same set of $i$-particles in
    parallel.}
  \label{fig:accGPU}
\end{figure}

We use predicted positions and velocities of the $j$-particles to
calculate gravitational forces that the $j$-particles exert on the
$i$-particles.  We schematically depict the parallelisation strategy
for such problem in Fig. \ref{fig:accGPU}. As before, we assume that
the total number of the $j$-particles is equal to $n$ and the total
number of the $i$-particles is equal to $m$.

We split the problem in $P$ parallel blocks, where $P$ is the number
of parallel multiprocessors (MP) on the GPU. The $j$-particles are
distributed evenly among the $P$ MPs. Each of the MPs then computes
the partial gravitational forces, exerted by the $n/P$ $j$-particles
assigned to that MPs, on all the $i$-particles in parallel.  This is
accomplished by assigning each of the $m$ parallel threads one of the
$i$-particles in the block. Here, we assume that the number of
$i$-particles $m$, is smaller than or equal to the maximal number of
parallel threads that a block is able to execute. Should $m$ be
larger, we split the $i$-particles in segments such that each of the
segments contains the maximal number of $i$-particles that can be
processed in parallel; the segments themselves are processed in a
serial manner.

The $P$ parallel blocks have no means of communication with each
other, whereas the threads within a multiprocessor are able to
exchange information via shared memory, which can be considered an
equivalent of an on-chip cache memory. Therefore, once all the
$i$-particles are processed, partial accelerations from each of the
blocks must be accumulated.

\begin{figure}
  \begin{center}
    \includegraphics[scale=0.75]{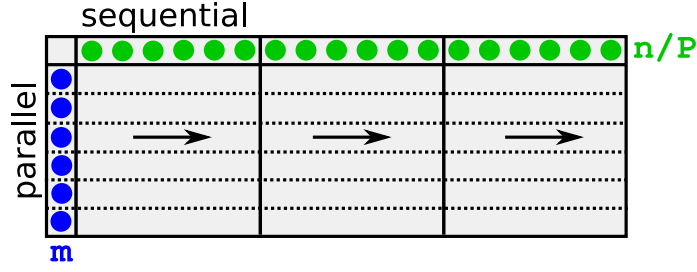}
  \end{center}
  \caption{Illustration of the decomposition on one multiprocessor. 1)
    Each thread reads in parallel one of the $n/P$ particles to shared
    memory; the total number of shared-particles is therefore equals
    to $m$. 2) These shared particles are processed sequentially.
    Steps 1) and 2) are repeated until all the $n/P$ particles have
    been processed.}
  \label{fig:accMP}
\end{figure}

The parallelisation method that we use in each block is illustrated in
Fig. \ref{fig:accMP}. Each thread in a block loads one particle from
the global memory to the shared memory, and therefore the total number
of particles stored in shared memory (shared particles) equals to the
number of threads in a block. Afterwards, each thread sequentially
calculates and sums the partial gravitational forces from the shared
particles. Once completed, this loop is repeated until all of
$n/P$-particles in the block have been processed.

\begin{figure}
  \begin{center}
    \includegraphics[scale=0.75]{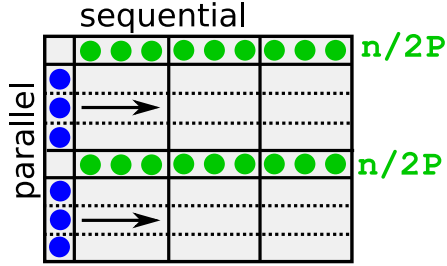}
  \end{center}
  \caption{Illustration of the decomposition on a multiprocessor in
    the case when the number of $i$-particles is equal to half the
    number of parallel threads that can be executed simultaneously. In
    this case, every particle is processed by two threads in order to
    guarantee that all threads are active. However, every thread
    processes half of the $j$-particles that are associated with this
    MP.
    \label{fig:accMP1}
  }
\end{figure}

If the number of the $i$-particles is smaller than the maximal number
of threads that a block can execute, it is desirable to further
parallelise the algorithm; we show this schematically in
Fig. \ref{fig:accMP1}. Here, we assume that the number of the
$i$-particles equals to half the number of threads that can be
executed in parallel. Thus, we split the $n/P$ $j$-particles in two
equal parts, such that each thread processes $n/(2P)$ particles. In
this way, the load per thread decreases by a factor of two, but since
the total number of the $i$-particles that are processed concurrently
doubles, the performance is unaffected. If we were to refrain from
this parallelisation step, the performance would suffer since there
would not be enough $i$-particles to fully occupy the multiprocessor,
and half of the parallel threads would remain idle.

The final step is to sum up the partial forces computed by each of the
$P$ parallel blocks.  A simple approach, in which all the data from
the GPU memory is first copied to the host and then reduced using the
CPU, is sufficiently efficient \citep{2008PHD_NITADORI}. Here, we have
chosen to reduce the partial forces directly in the GPU memory and
only copy back the final result.

The gravitational accelerations and jerks are sensitive to a round-off
errors which occur during the calculation of distances between two
particles.  We decrease these round-off errors by calculating this
distances using DS arithmetics. This can be understood considering the
following example: if the first four significant digits in the
position of the two particles are the same and the positions are
stored in SP, the resulting distance has only three significant
digits. The use of such low accuracy result degrades the quality of
the accelerations and jerks, and the accuracy of the simulation
overall. If, however, the positions are stored in DS, the number of
significant digits in the distance is nine instead of three, and this
is sufficient for an accurate integration. Since the rest of the
calculations is carried out in SP, the significant digits beyond the
seventh can be discarded, implying that the separation, even though
computed in DS, can be stored in SP.

The actual implementation is quite simple. One may bluntly apply
formulae to calculate the difference between two DS numbers. Such
method is expensive as it requires approximately ten SP floating point
operations to subtract two DS. However, such operation takes proper
care of both the most and the least significant digits in the
result. However, for our purpose, we can simplify the subtraction by
discarding the least significant digits. In this way, we reduce the
number of FLOPs. If the positions of two particles are stored in DS
format, ${\bf x}_i = \{{\bf x}_{i,{\rm hi}}; {\bf x}_{i,{\rm lo}}\}$
and ${\bf x}_j = \{{\bf x}_{j,{\rm hi}}; {\bf x}_{j,{\rm lo}}\}$, the
separation $\Delta{\bf x}_{ij}$ is
\begin{equation}
  \Delta{\bf x}_{ij} = 
  ({\bf x}_{j,{\rm hi}} - {\bf x}_{i, {\rm hi}}) + 
  ({\bf x}_{j,{\rm lo}} - {\bf x}_{i, {\rm lo}}).
  \label{eq:xij}
\end{equation}
The number of FLOPs required to carry out this difference is equal to
3. With this implementation, we will be able to resolve particles
separations to one part in ten millions or better.

This is the only operation that is carried out in DS
arithmetics. Since it is usually assumed that it takes 60 FLOPs to
calculate gravitational accelerations and jerks, the number of FLOPs
that our method requires is therefore equals to 66. Here, we
substitute each of the three SP subtraction operations with
Eq.\ref{eq:xij}.

\subsection{Parallelization over multiple GPUs}\label{Methods:GPUParallel}

Our library has a built-in support for multiple GPUs installed on a
single host PC. The support of multiple GPUs is done with the help of
the {\tt GPUWorker} library which is a part of the {\tt HOOMD}
molecular dynamics GPU-code \citep{Anderson20085342}. The
parallelisation is carried out automatically, based on the
availability of multiple GPUs and the user request given in a
configuration file. The application that uses the library is unaware
of this, meaning that no modifications to source code are required for
using multiple GPUs on one host. The parallelisation strategy in {\tt
  Sapporo} is rather straightforward: given a number $Q$ of GPUs, the
library distributes $n$ of the $j$-particles equally between all of
the GPUs. As a result, each of the GPUs processes $n/Q$ of
$j$-particles, but the same set of $i$-particles. Using multiple GPUs
results in a speed up for sufficiently large $n$, otherwise the
smaller number of $j$-particles per GPU may result in a performance
loss.

\subsection{Interface}

We designed the library with the same application interface (API) as
the standard GRAPE6 library.  To replicate the GRAPE-6 functionality,
{\tt Sapporo} supports softening per block of $i$-particles and
searches for nearest neighbours for each of the $i$-particles. This
makes {\tt Sapporo} suitable for high-precision collisional $N$-body
simulations. The swapping of libraries during the linking process
allows existing applications which already support the GRAPE6 API to
be used directly with {\tt Sapporo} without any changes to the
software\footnote{The number of pipelines in {\tt Sapporo} is $256$
  instead of $48$ for the GRAPE6, which may require resizing some
  arrays if an $N$-body code is written in a programming language with
  static memory allocation.}.  Compilation of {\tt Sapporo} does
require the presence of a CUDA-enabled GPU on the host computer as
well as the CUDA run-time libraries. From our experience,
GRAPE6-enabled $N$-body codes, such as {\tt Starlab}
\citep{2001MNRAS.321..199P}, {\tt phiGRAPE} or {\tt
  phiGRAPEch} \citep{2008MNRAS.389....2H}, and {\tt NBODY4}
\citep{1999PASP..111.1333A} can be used with the {\tt Sapporo}
library without any modifications to their source code.

 \section{Results}\label{sect:Results}

We measure the performance of {\tt Sapporo} by integrating equal mass
Plummer \citep{1915MNRAS..76..107P} spheres with a various number of
particles, $N$, for a fraction of an $N$-body time unit
\citep{1986LNP...267..233H}\footnote{\tt
  http://en.wikipedia.org/wiki/Natural\_units\#N-body\_units}. These
experiments were performed using two different codes for high
precision collisional $N$-body simulations, {\tt Starlab}
\citep{2001MNRAS.321..199P} and {\tt phiGRAPE}
\citep{2007NewA...12..357H}. We also performed an integration of a
$N=32k$ Plummer sphere well beyond core collapse.

The host computer used for these tests is equipped with one Intel
Core2 Quad CPU operating at 2.5GHz, 2GB of RAM, and two NVIDIA GeForce
9800GX2 graphics cards. Each of these cards is equipped with two
independent G92 GPU chips. For the user the system appears to be
equipped with four independent GPUs. We used the NVIDIA CUDA driver
v177.67, CUDA Toolkit v1.1 and CUDA SDK v1.1 and the installed
operating system is Debian GNU/Linux 4.0 with the 2.6.21 SMP x86\_64
kernel. The library is also compatible with CUDA Toolkit v2.1.

For comparison, some of the calculations were repeated on GRAPE-6A, for
which we used eight nodes of the Rochester Institute of Technology
cluster \citep{2007NewA...12..357H}. Each of these nodes is equipped
with a GRAPE-6A.  The code {\tt phiGRAPE} runs in parallel over several
nodes on this cluster, whereas in the GPU setup several GPUs are
hosted by a single PC. Some difference in performance can therefore be
attributed to the inter-node communication on the GRAPE cluster, which
is negligible on the single GPU-equipped PC. For comparing the
accuracy, we have also done a number of calculations on a single PC
with GRAPE-6A, which we refer to as GRAPE PC.  This PC is equipped with
a 2.8GHz Pentium D processor ewith 1GB RAM and a GRAPE-6A PCI card, and
running Debian 4.0 GNU/Linux with the 2.6.18 SMP x86\_64 kernel.

\subsection{Performance}

\begin{figure}
  \begin{center}
    \includegraphics[scale=0.5]{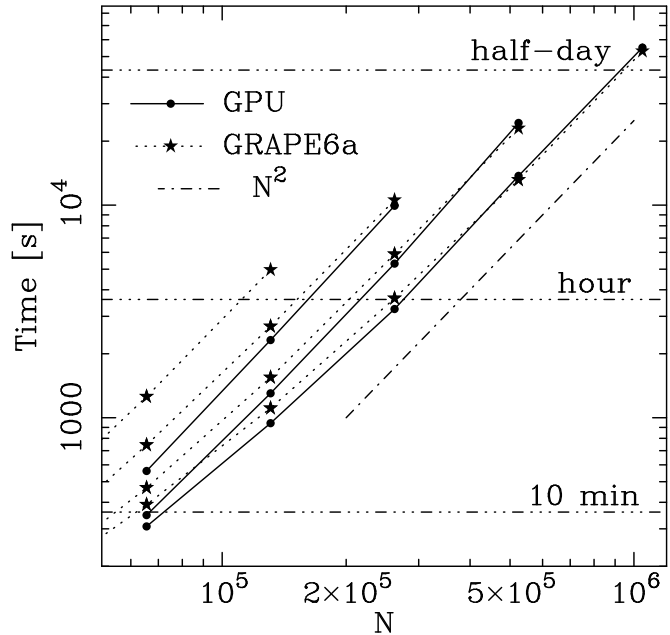}
  \end{center}
  \caption{The wall-clock time as function of the number of particles
    using GPUs (solid line with bullets) or GRAPE-6A (dotted line with
    stars). The bullets and stars are measured from integrating a
    Plummer models with $N$ particles using {\tt phiGRAPE}. From top
    to bottom, one, two, or four GPUs and one, two, four and eight
    GRAPEs were used. The dashed-dotted line shows the expected
    $N^2$-scaling, offset not to overlap with the measurements.}
  \label{fig:wc}
\end{figure}

The results of the performance tests are presented in
Fig.~\ref{fig:wc}, where we show the wall-clock time as a function of
the number of particles, $N$, for 1, 2 and 4 GPUs.  In all the cases,
the wall-clock time scales as expected, with $N^2$ and the wall-clock
time decreases as the number of GPUs increases. Simulations with $N
\le 64k$ on 4 GPUs show a degradation in the speed-up, which is caused
by the smaller number of particles in each block per GPU, see
\S\,\ref{Methods:GPUParallel}, and we notice similar behaviour for the
GRAPE-6A cluster \citep{2007NewA...12..357H}. The performance of two
GRAPE-6A-nodes is comparable to that of a single GPU.

\begin{figure}
  \begin{center}
    \includegraphics[scale=0.5]{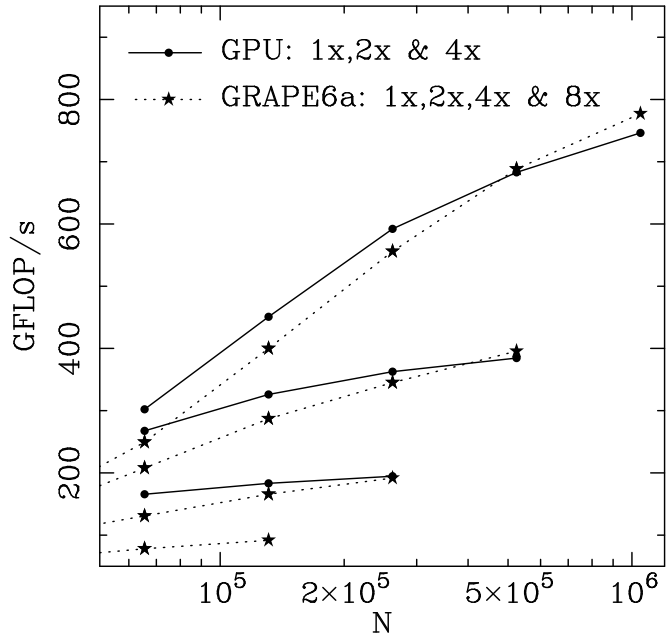}
  \end{center}
  \caption{The computational speed (in GFlops) as a function of $N$
    using GPUs (solid lines with bullets) and GRAPE-6A (dotted lines
    with stars) using the same initial conditions as using to generate
    Fig.\,\ref{fig:wc}. The number of GPUs and GRAPEs used for these
    measurements increases from bottom to top.}
  \label{fig:flops}
\end{figure}

In Fig.~\ref{fig:flops} we show the sustained performance in
GFlop/s\footnote{The exact number of floating point operations
  required for one force calculation varies in the literature. Here we
  assumed that a single force calculation requires 60 operations,
  which was also used for determining the theoretical peak performance
  of GRAPE \citep{2005PASJ...57.1009F}} for a range of $N$.  The
performance of the GPU is better than that of the GRAPE-cluster by
about a factor of two.  The relatively slow PCI-bus used in the GRAPE
nodes causes the performance for very small $N$ to be even worse
compared to the GPU-equipped PC.

The average number of particles in a block time-step can also be used
as a diagnostic tool. In {\tt phiGRAPE} the block size is controlled
by the standard time step criterion \citep{1985mts..conf..377A}. This
criterion includes a term, that depends on the difference in force on
a the particle at the beginning and end of a time step. This can have
an effect on time steps (and therefore block size) if the forces are
calculated with low precision. We found that small time steps tend to
become much smaller in this case, as the relative error in the force
difference becomes larger (when time steps are small, the force is
usually large but it changes very little over this time step,
resulting in the same problem as discussed before for calculating the
distance of two particles). Smaller time steps generally result in
smaller blocksizes, which in turn results in an inefficient use of the
GPU and also increases the number of integration steps needed.

\begin{figure}
  \begin{center}
    \includegraphics[scale=0.5]{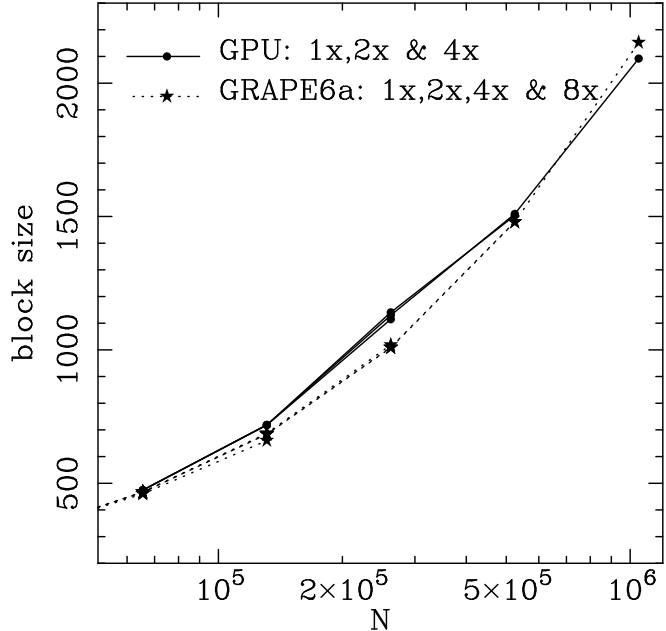}
  \end{center}
  \caption{The average number of particles in a block as a function of
    $N$.}
  \label{fig:block}
\end{figure}

In Fig.~\ref{fig:block} we present the average block size as a
function of $N$ for GRAPE-6 as well as for GPU.  The difference between
the block size for GRAPE and GPU is small for the adopted standard
time-step criterion, which indicates that the accuracy in the force
calculation on the GPU is comparable to that of GRAPE. It is important
to note, that the emulation of double precision is the key for
achieving this accuracy.

\subsection{Accuracy}

The accuracy of the force evaluation and integration of the equations
of motion is crucial for a reliable $N$-body simulation. We therefore
test in this section the adopted method for achieving high accuracy on
the GPU hardware and compare this with GRAPE-6.  There are various
sources of error in any $N$-body integration: the most straightforward
comes from the integration scheme itself and depends on the order of
the integrator; another error is caused by the limited (single)
precision of the GPU. The error in the integration is controlled via
the accuracy parameter $\eta$ \citep{1985mts..conf..377A}, such that
smaller $\eta$ results in a lower energy error. For a sufficiently
small $\eta$\, the integration errors are dominated by the limited
precision of the hardware.

\begin{figure}
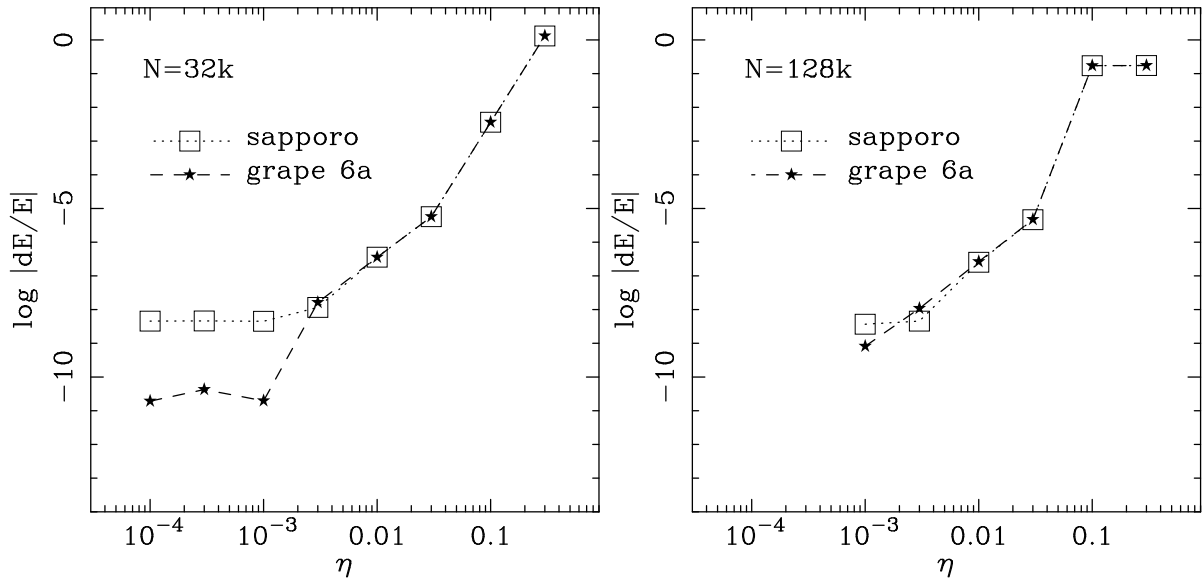

  \begin{center}
    \includegraphics[scale=0.45]{plots/err032k.ps}\vbox{}
    \includegraphics[scale=0.45]{plots/err128k.ps}
  \end{center}
  \caption{The relative error in total energy as a function of the
    time-step accuracy parameter $\eta$ for a Plummer model with
    $32$k (left panel) and $128$k (right panel) particles ware
    integrated for 1/4 $N$-body time units using {\tt phiGRAPE}. The
    dotted line with open squares shows the energy error for the GPU
    with {\tt Sapporo}, the dashed line are from GRAPE-6A.}
  \label{fig:devseta}
\end{figure}

To test accuracy of {\tt Sapporo} we integrated an equal mass Plummer
models with various $N$. For every $N$ we studied the dependence of
the relative energy error, $dE/E = (E_1 - E_0)/E_0$, as a function of
the accuracy parameter, $\eta$, and the number of particles, $N$.  For
this purpose, we varied $\eta$ and $N$ between $10^{-4}$ to $0.3$ and
16k to 256k respectively. Our results show that the error is strongly
dependent on $\eta$, whereas dependence on $N$ is rather weak. in
Fig.~\ref{fig:devseta} we show the results for $N=32$k and $N=128$k.

For $\eta \gtrsim 3\cdot 10^{-3}$, the relative errors $dE/E$
decreases with $\eta$ and the difference between GRAPE and GPU is
indistinguishable independent of $N$ (see
Fig.~\ref{fig:devseta}). This behaviour is expected, since for
relatively high $\eta$ the energy error is dominated by the
integration errors rather than by the errors in the forces evaluation.
For $\eta \aplt 3\cdot 10^{-3}$, however, the integration error
saturates at the point where the hardware limits the accuracy of the
force evaluation.  As expected for $\eta \aplt 3\cdot 10^{-3}$ the
error $dE/E$ does not depend on $\eta$, although the saturation level
is weakly dependent on $N$. On the GPU, the relative error is $dE/E
\simeq 10^{-9}$ due to the limited precision, which is of order of the
minimal relative error that can be achieved with single precision. The
errors produced by the GRAPE-6 saturates at a somewhat lower level fo
$dE/E \simeq 10^{-11}$, which is expected based on the higher accuracy
of the internal GRAPE hardware \citep{1998sssp.book.....M}. Note, that
typical values for $\eta$ are between $10^{-2}$ and $10^{-1}$, and
therefore well within the range where {\tt Sapporo} is not limited by
the single precision accuracy of the GPU.

\subsection{Star cluster evolution and core collapse}

\begin{figure}
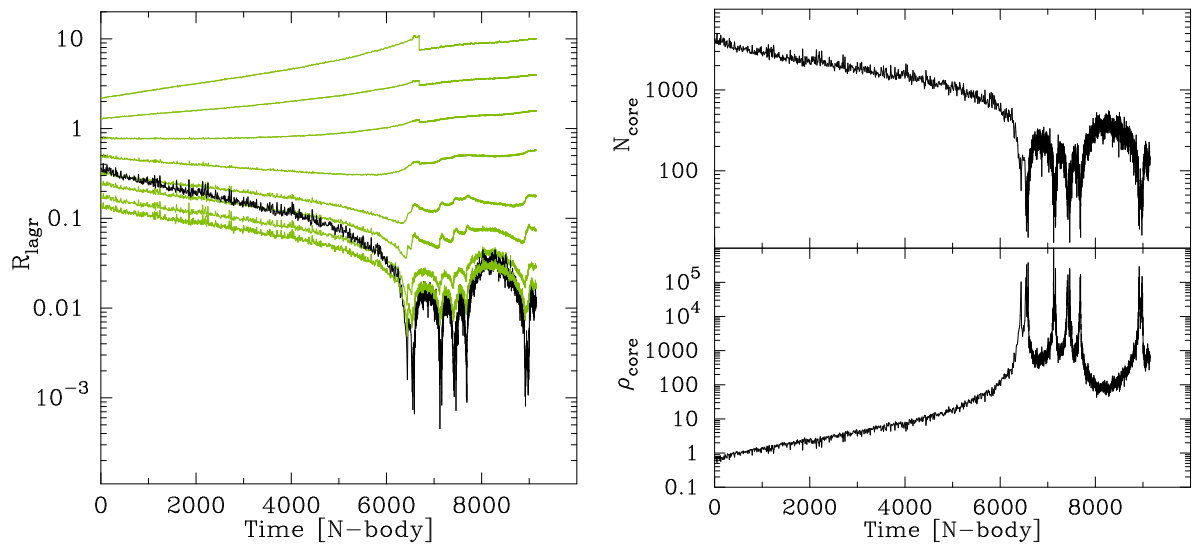

  \begin{center}
    \includegraphics[scale=0.4]{plots/rlagr.ps}$\quad$
    \includegraphics[scale=0.4]{plots/ncore.ps}
  \end{center}
  \caption{The left panel shows time evolution of the core radius
    (black line) and 1, 2, 5, 10, 25, 50, 75, 90\% lagrangian radii
    (green lines, from bottom to top). The right panel shows the time
    evolution of both the number of stars in the core (top) and core
    density (bottom). The initial condition was an equal-mass Plummer
    distribution with $N=32768$ particles. The core collapse is
    reached at $t\simeq 6440$. The calculation was terminated at $t=
    9200$. The average sustained speed for the entire calculation was
    about 113\,Gflop/s. The sudden decrease of largrangian radii at $t
    \simeq 6800$ is caused by the removal of the escaped stars.}
  \label{fig:rcore}
\end{figure}

We tested {\tt Sapporo} in a production environment by calculating the
time evolution of an equal-mass Plummer model with $N$=32k particles
through core collapse. The integration was performed using {\tt
  Starlab} $N$-body simulation environment, and we used only one GPU
for this calculation.

The results of this calculation are presented in Fig.~\ref{fig:rcore},
where we show time evolution of some cluster parameters. In
particular, it is evident that the cluster reaches core collapse at $t
\simeq 6440$ $N$-body time units. The wall-clock time it took to reach
core collapse was about 24 days, whereas it took the simulation only 9
days to reach the point half-way to core collapse. The whole
simulation, which we terminated at $t\simeq 9200$ $N$-body time units,
was completed in about $41$ days. We note, that a relatively
inexpensive commercial GPU, which we used for this experiment, proved
to be stable during the entire simulation.

The sustained performance up to $t \simeq 3000$\,$N$-body time units
was about 140\,Gflop/s, which reduced to about 132 Gflop/s by the time
core collapse was reached, and to drop even further to about
88\,Gflop/s in the post core collapse phase.  The lower performance
after core collapse is mainly caused by the presence of a dense core
and hard binaries, which formed during core collapse and which require
additional calculations performed on the host.  The mean sustained
performance for the entire simulation was about 113\,Gflop/s.

\section{Summary}\label{sect:discussion}

The integration of a large self gravitating system for many relaxation
times is a computationally demanding procedure. Currently, the GRAPE6
special-purpose hardware is the state-of-the-art for carrying out such
simulations. In this paper, we have presented a library named {\tt
  Sapporo}, which emulates the GRAPE6 functionality on NVIDIA GPUs
using the CUDA programming environment.  Our library is publicly
available and we emulate GRAPE6 interface as closely as possible (with
the exception of some rarely used functions). The library is designed
in such a way that it can be simply swapped with the original GRAPE6
library and, as a result, a wide range of current $N$-body programs,
that have been written using the GRAPE6 interface, can now be used on
GPUs using {\tt Sapporo} without any modification of the source
code. 

We have carried out a number of basic $N$-body experiments in order to
test the efficiency and accuracy of the library and we compared our
results with the GRAPE6. We found that our library is twice as fast as
commonly used GRAPE6A/GRAPE6-BLX cards, and it is as accurate as GRAPE
with standard integration parameters. In particular, the performance
of a PC equipped with two NVIDIA GeForce 9800GX2 cards is on par with
that of a 8-node GRAPE6A cluster or a 32-chip GRAPE6 board.

We have also carried out a production run to test the ability of the
library to handle a real astrophysical problem. For such stress-test,
we have chosen to integrate an equal-mass Plummer model beyond core
collapse. As expected, the cluster reaches core collapse in about 15
initial half-mass relaxation times and afterwards enters the phase of
gravothermal oscillations. Our results are consistent with those
previously published in the literature
\citep{2003gmbp.book.....H}. The ability of {\tt Sapporo} to handle
such a demanding problem demonstrates that the library can be safely
used for realistic astrophysical problems.

The high performance of the {\tt Sapporo} library is achieved by
splitting the steps of the Hermite integration scheme between CPU and
GPU in the same way as it is done when using the GRAPE6. In
particular, the prediction and calculation steps are carried out on
the GPU, whereas the corrector is computed on the CPU. In this way,
the force calculation is always a dominant part of the integration. If
the prediction for all particles is carried out on the host, as it is
done in the {\tt kirin} library \citep{2008NewA...13..103B}, the
integration would be dominated by the prediction step and
communication, and thus degrading the performance.

It is common nowadays, that a single production node is equipped with
at least two GPUs. Therefore, we designed the library to make an
efficient use of such a configuration, and the user can configure the
library to use any combination of available GPUs. The application
which makes use of the library is unaware that multiple GPUs are
utilised, i.e. no modification to its source code is required.

We also implemented a neighbour lists functionality in {\tt
  Sapporo}. The maximum number of neighbours that a particle in a
block can store is configured at compile time (default is 256), and it
cannot be changed during runtime. Calculating and storing neighbour
lists has an impact on performance of a few tens of percent, depending
on the maximum number of neighbours allowed. On the other hand, it is
also possible to disable neighbour lists all together, and by this
gain up to 50\% in speed.

An important part of {\tt Sapporo} is the emulation of double
precision using single precision numbers by using standard methods
from the literature. Similar techniques were implemented in the {\tt
  Yebisu} $N$-body code which also runs on NVIDIA GPUs
\citep{2008PHD_NITADORI}. The partial emulation of double precision
has a minimal impact on performance, and places current and future
NVIDIA GPUs on a very competitive ground with current and future GRAPE
hardware.

The emulation of double precision was required because the previous
generation of NVIDIA GPUs (based on G80/G92 architecture) does not
support it. Current NVIDIA GPUs (based on GT200 architecture) have
native double precision support, however, only for a limited number of
pipelines. This enables in principle the use of double precision, but
the performance is an order of magnitude smaller in that
case\footnote{\tt http://www.nvidia.com/object/product\_tesla\_s1070\_us.html}. The
GT200 architecture also has almost three times as many parallel
elements and twice the theoretical peak-performance of the previous
G80/G92 architectures. Using a GT200 GPU, we were able to get a
noticeable speedup of about 50\%, which is smaller than the factor of
two in peak performance. The reason is, that {\tt Sapporo} was
designed for the older GPU architectures, and therefore does not in
the current version efficiently utilise all the parallel elements of a
GT200 GPU. However, we are currently working on a new version of {\tt
  Sapporo} which will be scalable across future generations of NVIDIA
GPUs, including the GT200.

\section*{Acknowledgements}

We thank Keigo Nitadori, Jun Makino and Steve McMillan for valuable
suggestions and help in developing the library. This work was
supported by NWO (grants \#635.000.303, \#643.200.503).  This research
was supported in part by the Netherlands Research School for Astronomy
(NOVA), the National Science Foundation under Grant No. PHY05-51164,
and the KITP in Santa Barbara for their hospitality.  We are grateful
to Drexel University (Steve McMillan), Rochester Institute of
Technology (David Merritt) and European Southern Observatory (Mark
Gieles and Markus Kissler-Patig) for the use of their GRAPE-6 hardware,
and to NVIDIA (David Luebke) for the GPU's used for this paper.

\bibliographystyle{elsarticle-harv} 
\bibliography{GHPZ08sapporo}

\end{document}